\documentclass[12pt]{JHEP3}

\usepackage{amsmath,amssymb,epsf,epsfig}

\newcommand{\Tr}{{\rm Tr\,}}

\newcommand{\diag}{{\rm diag\,}}
\newcommand{\Z}{{\bf Z}}

\preprint{SNUTP 05-009}
\title{Gauge Unification\\ via Stable Brane Recombination}
\author{Kang-Sin Choi and Jihn E. Kim\\
 School of Physics and Center for Theoretical Physics, \\
 Seoul National University, Seoul 151-747, Korea \\
 E-mail: \email{ugha@phya.snu.ac.kr,\ jekim@phyp.snu.ac.kr}}

\abstract{We study the transition between parallel and intersecting
branes on a torus. Spontaneous symmetry breaking of nonabelian gauge
symmetry is understood as brane separation, and a more general
intermediate deformation is discussed. We argue that there exists
supersymmetry preserving transition and we can always have parallel
branes as a final state. The transition is interpreted due to
dynamics of the F- and D-string junctions and their generalization
to (F, D$p$) bound states. The gauge group and coupling unification
is achieved, also as a result of supersymmetry. From the tadpole
cancelation condition, we naturally have intersecting brane models
as broken phases of Type I theory with $SO(32)$ gauge group.}


\begin{document}

\section{Introduction}

Unification of gauge couplings is an appealing idea. In field
theory, it is neatly realized in grand unified theories \cite{GQW}
and also  arises naturally in compactification of heterotic string
theories \cite{HetUni}. In many models gauge couplings are unified
at the string scale \cite{IKNQ} (although the hypercharge
embedding in such models in general cannot give a desirable value
for the weak mixing angle, $\sin^2\theta_W^0 = \frac38$
\cite{BrSLM}).

In this paper, we show that the gauge group and coupling
unifications are still present in the intersecting brane models.
In Type I/II strings, the open strings describe gauge symmetry and
they end on D-branes. If $N$ branes are stacked parallel, we have
a $U(N)$ gauge symmetry out of which the $SU(N)$ coupling is the
unified one. Their matrix-valued brane positions $X^m$ are
translated into vacuum expectation values of gauge field $A'_m$
$$ X^m = 2 \pi \alpha' A^{\prime}_m, $$
in the $T$-dual space. Depending on the eigenvalues, we have
spontaneous symmetry breaking of gauge group, which is also
translated into brane separation. If we can make use of the above
idea of brane separation, we can understand unification. Some
considered such unification partially in the parallel stacks
\cite{CLLL} and even obtained some relations between couplings
\cite{BLS}.

If a stack of $N_1$ branes and a $N_2$ branes intersect at angle,
then the reduced four dimensional $U(N_1)$ and $U(N_2)$ couplings
are independent. They are thus different in general because of the
branes' different wrappings of internal world volumes.

We will show, in Section 3, that {\it transition between parallel
and intersecting branes}, while preserving supersymmetry, can be
achieved by turning on electric flux. Roughly speaking, we
consistently introduce a more general control on the set of
parameters describing the brane separation as
$$
 X^m = \diag(f_1(X),f_2(X),\dots,f_N(X)),
$$
where $f_i(X)$ describes a fluctuation of branes in $X^m$ direction
and have a general shape. Although a general setup breaks
supersymmetry, we recover supersymmetry by turning on electric flux,
whose source is nothing but the F-string. This has been understood as
string junction \cite{Sch,Wi,HT,CM,DM} and its higher dimensional
generalization. The key property is that any deformation does not
change the supersymmetry condiion or the BPS condition and the
process is reversible. We can obtain the final state as a stack of
(coincident or non-coincident) parallel branes.

To understand such a setup, one notes the crucial interplay
between F- and D-strings. The supersymmetry or BPS condition is
interpreted as the balance condition at the junction they form.
Their descriptions are similar except that their tensions differ
by the factor of string coupling. In the $T$-dual picture,
F-strings and D-branes become sources for electric and magnetic
fluxes, respectively. Therefore, not only the D-brane but also the
F-string dynamics are translated and unified into those of gauge
fields. We will see that these together affect the four
dimensional Yang--Mills coupling constant.

In Section 4, when we consider the Ramond--Ramond (R--R) tadpole
cancelation condition, which is the consistency condition of open
strings, we observe that {\it some class of models is naturally
converted into some $T$-dual version of Type I string theory with
gauge group $SO(32)$}. We can reinterpret this result as the
intersecting brane models being obtainable by compactifying Type I
string with $SO(32)$.

\section{Intersecting branes}

\subsection{Brane separation and symmetry breaking} \label{branesep}

When there are $N$ sheets of coincident D$p$-branes, we have
$U(N)$ gauge fields \cite{PoWi}. Described by the Chan--Paton
factor, a charge is attached on either end of an open string. An
open string ending on branes transforms as the adjoint
representation $\bf N^2$ of $U(N)$, because the open string
between different branes behaves like a charged boson. It is shown
that the lowest lying degrees along the brane are gauge fields and
a transverse fluctuation is described by a scalar field. We have
$T$-duality relations between them and the Dirac--Born--Infeld
(DBI) action
\begin{equation} \label{DBI}
 S = - T_p \int d^{p+1} x \Tr e^{-\Phi}
  \sqrt{-\det(G_{\mu \nu}+B_{\mu \nu}+2\pi \alpha' F_{\mu \nu})}
\end{equation}
takes account of them. Expanding to the quadratic order in
$\alpha' F_{\mu \nu}$, it reduces to ($p$+1)-dimensional Yang--Mills
(YM) action,
\begin{equation} \label{derivedYM}
 S_p = - {T_p (2\pi \alpha')^2 \over 4 g_{\rm s}}
  \int d^{p+1} x ~\Tr F_{\mu \nu} F^{\mu \nu}
\end{equation}
with a potential of transverse scalar degrees. Here $T_p$ and
$g_{\rm s}$ are the tension and Type II string coupling fixed by
the vacuum expectation value (VEV) of dilaton. Therefore, the YM
coupling is
\begin{equation}
 g^2_{p+1} =\frac12 g_{\rm s} T^{-1}_p (2\pi \alpha')^{-2}.
\end{equation}

Location of branes $X^m$, which is matrix valued, is translated
into gauge field $A_m^{\prime}$ in the $T$-dual space
\begin{equation} \label{wilson}
 X^m = 2 \pi \alpha' A'_m\,
\end{equation}
where  $A'_m = \sum A_m^{\prime a} T^a$. We put prime to indicate
that the gauge field is in the $T$-dual space.  We will use Greek
indices for the worldvolume directions and lower case Roman
indices for the transverse directions to the branes.  The constant
field can be always diagonalized by a suitable gauge
transformation to
\begin{equation} \label{diagwilson}
 X^m = \diag(a_1,a_2,\dots,a_N).
\end{equation}
From geometry, one can investigate the group structure. It
corresponds to separation of branes located at $X^m =
a_1,a_2,\dots,a_N$. At generic values of $a_i$s, $U(N)$ breaks down
to $U(1)^N$. Some of eigenvalues may be coincident, whose number we
call  $N_k$ where their sum is $N= \sum N_k$, so that the branes
form separated stacks. It results in breaking $U(N)$ down to
$U(N_1)\times U(N_2) \times \cdots \times U(N_k)$. The $U(1)$ factor
in $U(N_i)$ corresponds to the overall translational symmetry of
each stack. The transverse directions to the brane are compactified
on small radii; thus the components $A'_m$, in turn $X^m$, are
decoupled and become scalar fields. It is understood as a Higgs
mechanism that the adjoint Higgs $X^m$ develops a VEV. The potential
is proportional to $\Tr[X^m,X^n]^2$, which comes from the T-dual of
$(\partial_{[m} A_{n]})^2$, and diagonalized matrix
(\ref{diagwilson}) gives flat directions for any values of $a_i$,
thus supersymmetry is preserved at all the vacua. $X^m$ is the
order parameter of symmetry breaking, which will be generalized to
a more complicated setup.

\subsection{Intersecting brane world}

The above scalars form a real adjoint representation. To have a
realistic model, we need chiral fermions. They naturally emerge
from intersecting branes \cite{BDL}. When we have $N_a$ and $N_b$
coincident branes intersecting with an angle, we obtain chiral
fermions transforming as a bi-fundamental representation $\bf
(N_a, N_b)$ under $U(N_a) \times U(N_b)$ (or $\bf (N_a,
\overline{N}_b)$, if instead we placed $N_b$ anti-branes). They
are localized at the intersection, to minimize the energy
proportional to the stretched string length.

\begin{figure}[t]
\begin{center}
\epsfig{file=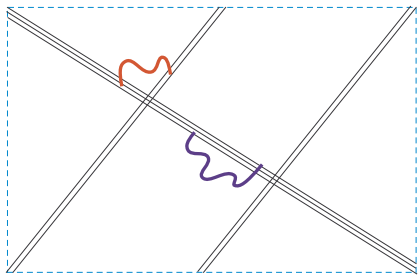,height=4 cm}
\end{center}
\caption{ Intersecting branes with two families of $\bf (3,2)$
quarks charged under $SU(3)\times SU(2)$.} \label{intbrane}
\end{figure}

In a typical setup \cite{IMR,Ur,BCLS} we have compact dimensions and
each stack of branes wraps a cycle on it, and furthermore we have
intersections between brane stacks in general. Fig. \ref{intbrane}
shows a typical setup showing the basic features of intersecting
branes. We have two and three slices of branes wrapping one-cycles
in the two-torus, which account for $U(2)$ and $U(3)$ gauge groups,
respectively. There are a finite number of intersection points in
general. The same copy of intersections naturally explains the
number of families and Yukawa couplings in the low energy theory are
obtained geometrically \cite{CIMYu,HV,FINQ}. The standard setup is a
D6-brane wrapping the six-torus in Type IIA theory branes. Out of
six, three internal dimensions wrap one-cycle in each two-torus and
the remaining (3+1) dimension takes account of our four dimensional
world.

The typical intersecting brane world models suffer the following
problems:  the problem of gauge coupling unification, and the
instability problem in non-supersymmetric models:
\begin{itemize}
\item[(i)]
The semisimple gauge group is obtained  from the different stacks of
branes in the construction, and hence the disconnected parts are not
related to each other. Therefore, the gauge couplings are
independent. Consider the dimensional reduction from the ($p$+1)
dimension to (3+1) dimension
\begin{equation} \label{dimredux}
 -{ 1 \over 4g_{p+1}^2} \int d^{p+1}x |F_{\mu \nu}|^2 \quad
 \longrightarrow \quad -{V_{p-3} \over 4g_{p+1}^2}  \int d^4x
 |F_{\mu \nu}|^2.
\end{equation}
Therefore, the four dimensional coupling constant squared is
inversely proportional to the compact volume $V_{p-3}$ that each
brane cycle wraps
\begin{equation} \label{4dcoupling}
 g_4 ^2 = {g_{p+1}^2 \over V_{p-3}}.
\end{equation}

For the coupling constant unification, there is a delicate issue
on the non-standard normalization of $U(1)$ couplings.
\item[(ii)]
When we perform mode expansions of intersecting branes with
suitable boundary conditions, we instantly see that the lowest
excitation is the tachyon for a generic angle $\theta$ between
branes,
\begin{equation}
 m^2 = - {\theta \over 2 \pi \alpha'}, \quad 0 \le \theta
 \le \pi.
\end{equation}
For more than one dimensions, $\theta = \sum (\pm 1) \theta_i $ is
the sum of angles of the sub-tori $T^2$. This signals the brane
instability \cite{HN}, and the expansion is  around the wrong
vacuum. It is hoped that string field theory takes care of this
instability problem. In this context, the tachyon condensation has
been studied \cite{Se}.
\begin{figure}[t]
\begin{center}
\epsfig{file=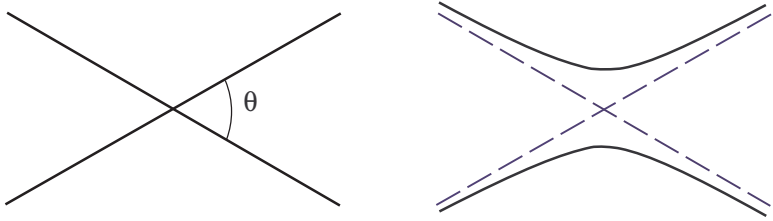,height=3.5 cm}
\end{center}
\caption{Without supersymmetry the intersecting branes are
unstable. There is a recombined state that has a lower energy than
that of the intersecting branes and decays to this lower energy
state. The decay process goes on until the branes become flat.}
\label{recomb}
\end{figure}
The supersymmetry route can be taken, where one can find some
combinations of angles vanishing in higher dimensions for the
lowest energy. Alternatively we might abandon supersymmetry and
this tachyon phenomenon is used as a Higgs mechanism \cite{CIMHi},
in which case the electroweak scale $M_Z$ (or the Higgs mass)
should be comparable with compactification scale $1/R$ (or the
separation of branes). However, from the point of view of
unification and supersymmetry, this scenario is not so favorable.
Another cure comes from {\it turning on electric or magnetic
field} to recover supersymmetry, which is our viewpoint taken in
the present paper.
\end{itemize}

We can understand the chirality of fermions as follows. Under the
breaking $U(N) \to U(N_a) \times U(N_b)$, the branching rule for
adjoint looks
\begin{equation} \label{branching}
 \bf  N^2 \to (N_a^2,1) + (1,N_b^2) + (N_a,N_b)
    + (\overline{N}_a, \overline{N}_b)
\end{equation}
where $N=N_a+N_b$. In the parallel brane case, the resulting
massless fields are only the adjoints of the unbroken groups $\bf
(N_a^2,1) + (1,N_b^2)$. However, in the intersecting brane case, the
would-be massive chiral fields $\bf (N_a,N_b)$, sometimes called
$X^{\pm},Y^{\pm}$ bosons in the $SU(5)$ GUT for example, become
massless at the intersecting point. Note that we should count the
fermion $\bf (N_a,N_b)$ only once because its complex conjugate
representation corresponds to its $CPT$ conjugate. This is the
typical situation in string theory, because the complete state is
formed with the right movers which provide the chirality
\cite{BDL,Ch}. Note that the change of the metric (i.e. the complex
structure) does not change the presence of chiral fermions and the
spectrum, although it affects the nonvanishing masses of fermions.
This reveals some structure that such setting can be {\em embedded
into the bigger group $U(N)$.}

\subsection{Tilted but parallel branes}

Consider a rectangular two-torus $T^2$ with sides $L_1$ and $L_2$.
We introduce the parametrization related to the winding number on
the homology cycle $[\pi_a] \in H_1(T,\Z)$. In two-torus we have two
basis $[a]$ and $[b]$ each being an equivalence class of curves
continuously shrinkable with the end points at each edges. We will
use the winding vector $(n_a,m_a) \equiv n_a [a] + m_a [b]$ which at
the same time denotes the R--R charge components of the brane in the
given torus. We restrict the discussion to the direct product of
two-tori, so that every cycle is represented by the direct product
of such vectors.

A single brane winding a cycle $(1,m)$ describes a U(1) gauge field.
The consistency requires that the brane should be {\it closed in the
compact dimensions, not end}. This tilted brane can be understood
also as the $T$-dual of a two dimensional brane with a constant flux
turned on \cite{Ba,BGKL,BT}. It is noted that the DBI action is the
unified description taking into account the both interpretations
\cite{Po,Ba}. In the $T$-dual space, we obtain $F_{21}= {2 \pi \over
L_1 L_2} m$ where $m$ is quantized, $m \in \Z$. For the
configuration of $U(1)$ bundle with the first Chern class $c_1=m$, it
is nothing but the monopole or vortex quantization condition.
Therefore, for such a tilted brane, there does not exist an extra
modulus describing the angle because the shape of torus and the
quantization condition totally determines the angle.

When we have a winding $(n,m)$, it cannot be described by an Abelian
gauge group even though it is a single D-brane. We are forced to
introduce the subgroup of the nonabelian gauge group
$U(n)=(U(1)\times SU(n))/\Z_n$ from two requirements: the shape of
torus and the quantization condition of the angle $\theta$. The
consistency condition requires the quantized flux $F_{21}= {2 \pi
\over L_1 L_2} \frac m n$ with $n,m \in \Z$, which has the same
first Chern class $c_1=m$. We can always choose the gauge \cite{AHT}
\begin{equation} \label{paranocoinc}
{1 \over 2 \pi \alpha'} X^2 = F_{21} X^1 + {2 \pi \over L_2} \diag
\left(0, \frac 1n, \dots, \frac{n-1}n \right),
\end{equation}
up to a constant offsetting the origin.  We have parallel branes
with slope $\tan \theta = F_{12}$ from the $X^1$ axis and equal spacing
$1/n$, which corresponds to $U(1)^n$ obtained from spontaneous
symmetry breaking of $U(n)$. We have parallel branes with slope
\begin{equation} \label{magslope}
 \tan \theta = {m \over n}.
\end{equation}
Of course each has the same gauge coupling because winding volume
in (\ref{4dcoupling}) is the same.

\subsection{Intersecting branes at angles}

 Even when branes are intersecting with an angle, we have a
similar order parameter description as done in
(\ref{paranocoinc}). It is more convenient to describe in the
$T$-dual space, using the dual gauge field $A_\mu$. Consider two
sheets of D2-branes extended in the $X^1$-$X^2$ direction and the
constant magnetic flux $F_{12}$ turned on it, which describes a
$U(2)$ gauge theory. By an appropriate gauge transformation, we
can always make the $F_{12}$ matrix diagonal. We may always choose
a gauge $A_1=0$ and
\begin{equation} \label{su2gauge}
 A_2 = X^1 F_{12} = {2 \pi \over L_1 L_2}
 \begin{pmatrix} p X^1 & 0 \\ 0 & -p X^1
 \end{pmatrix} .
\end{equation}
$T$-dualizing along the $X^2$ direction, we obtain
\begin{equation} \label{magflux}
 X^{\prime 2} = 2 \pi \alpha'  A_2 = {4 \pi^2 \alpha' \over L_1 L_2}
 \begin{pmatrix} p X^1 & 0 \\ 0 & -p X^1
 \end{pmatrix},
\end{equation}
which is linear in $X^1$. Seen componentwise, this indicates that we
have two pieces of D1-branes, having angles
\begin{equation}
 \theta = \pm \tan^{-1} p,
\end{equation}
relative to the $X^1$ axis. Here the magnetic flux $p$ is integrally
quantized from the consistency condition and it is interpreted as
the winding number in the dual picture.

We can generalize this idea to the case with more branes and more
complicated magnetic fluxes. For a larger gauge group $U(N)$ broken
to $U(N_1)\times U(N_2)\times\dots\times U(N_k)$, we can always
diagonalze and choose gauge such that $A_2=X^1 F_{12}$, as in
(\ref{su2gauge}),
\begin{equation} \label{sungauge}
 A_2 = {1 \over 2 \pi \alpha'} X^{\prime2}
  =  {X^1 \over 2\pi \alpha'} \diag (v_1, v_2,
  \dots,v_k)\
\end{equation}
up to constant offset. Corresponding to $U(N_i)$ for $i=1,2,\cdots$,
we have $N_i$  identical value of $v_i$. Eventually it is equivalent
to turning on a constant flux of $A_2$, which is Hermitian. After
removing the diagonal constant corresponding  the overall $U(1)$, we
can always diagonalize the traceless $A_2$ by a unitary
transformation. $v_i=m_i/N_i$ is quantized with the first Chern
class $m_i$.

Ref. \cite{HT} considered the possibility of turning on constant
off-diagonal components of gauge fields. When we diagonalize $X$,
the branes in this frame are in a form of reconnected ones, looking
like Fig. \ref{recomb}. This is a special case arising from the fact
that $X$s are not mere numbers but matrices so that they are not
commutative: the geometry is not clear but fuzzy. The off-diagonal
components, which are proportional to some generators in another
frame, are turned on by continuous Wilson lines. Because of the
$X^1$ dependent terms in (\ref{magflux}), the constant terms cannot
be simultaneously diagonalized. The constant components do not
change the BPS condition; thus the resulting theory is
supersymmetric.

\section{Supersymmetry preserving brane recombination}

\subsection{Brane junction}

The supersymmetry transformation for gaugino is
\begin{equation} \label{BPS}
 \delta \chi = \Gamma^{\mu \nu} F_{\mu \nu} \epsilon.
\end{equation}
Vanishing $\delta \chi$ corresponds to the BPS condition of
Yang--Mills theory and the number of surviving supersymmetries is
determined by that of the invariant spinor components.

\begin{figure}[t]
\begin{center}
\epsfig{file=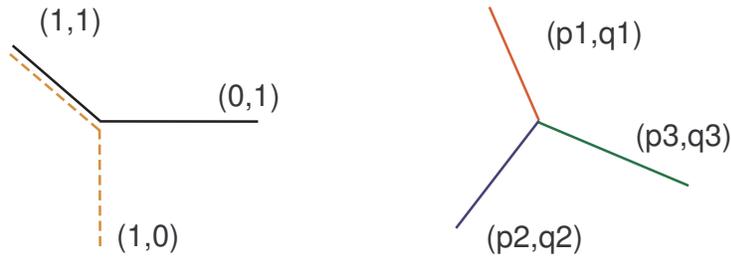,height=4 cm}
\end{center}
\caption{String junction. Its arbitrary deformation preserves the
supersymmetry as long as the junction condition is satisfied. This
is also understood as a (F, D$p$) bound state in higher
dimensions.} \label{junction}
\end{figure}

 Firstly, let us consider the simplest case of two dimensions
along the $X^1$-$X^2$ directions, where two D1-branes (or D-strings)
intersect at an angle. As seen before, in the $T$-dual picture,
the tilted brane is responsible for the field strength $2 \pi
\alpha' F_{12} = \partial_1 X^2$ and the winding numbers are
quantized. The branes being not parallel, there remains no
supersymmetry, because there is no common solution to
\begin{equation}
 \Gamma^{12} F_{12} \epsilon =0
\end{equation}
for two different $F_{12}$s. Nevertheless, we can reintroduce
supersymmetry by turning on an electric flux \cite{CM,DM}
\begin{equation} \label{D1reac}
 X^2 (X^1) = A_0 (X^1).
\end{equation}
Then, there is a solution of vanishing  (\ref{BPS}),
\begin{equation}
 (\Gamma^{12} F_{12} + \Gamma^{10} F_{10}) \epsilon =
 (\Gamma^{12} + \Gamma^{10}) F_{12}  \epsilon =0 \\
\end{equation}
or
\begin{equation} \label{survsusy}
 ( \Gamma^2 + \Gamma^0) \epsilon =0,
 \end{equation}
which has a nontrivial solution. Supersymmetry is recovered by the
component $F_{10}$ by compensating the back reaction of  D-string
deformation (\ref{D1reac}).  This source for electric flux is
interpreted as fundamental string (F-string), which has the
desired tension \cite{CM} .  Then the angle of the F-string about
the D-string is given by
\begin{equation} \label{chargeslope}
 \tan \theta = {p g_{\rm s} \over q}.
\end{equation}
 Notice that this expression contains string coupling constant
$g_{\rm s}$, thus is different from the setup with D-branes only
(\ref{magslope}). This is the condition of F1-D1 string bound
state (Eq. (13.6.3) of \cite{Po})
\begin{equation} \label{balance}
 T_{p,q} = \sqrt{ (p T_{1,0})^2 + (q T_{0,1})^2 },
\end{equation}
obtained from the supersymmetry algebra ((13.2.9) of \cite{Po}).
This is rewritten as
\begin{equation} \label{BPSnet}
 T_{p,q} \sin \theta = p T_{1,0}, \quad T_{p,q} \cos
 \theta = q T_{0,1} = \frac q g_{\rm s} T_{1,0},
\end{equation}
to reproduce (\ref{chargeslope}). From this, we observe that the
slope and the F and D charges are related as in Eqs.
(\ref{chargeslope}) and (\ref{BPSnet}). This is nothing but the
balance condition between the incoming and the outgoing branes,
since at every point
\begin{equation} \label{junccond}
 \sum p_i = 0, \quad \sum q_i = 0.
\end{equation}
It is also called the {\it string junction condition} shown in
Fig. \ref{junction}.
As long as this condition is satisfied, every string junction is a
BPS state preserving some supersymmetries allowed by
(\ref{survsusy}). It is natural to think of $p$ units of F-strings
and $q$ units of D-strings. When we compactify the spatial
dimensions on $T^2$, we require the charge quantization condition.
As seen in previous sections, we require $p$ and $q$ integers. We
do not identify $(p,q)$ as the winding number $(n,m)$ but the
identification is possible when we consider some small box
encompassed or totally filled  by a linear piece of the string.

In the balance condition between F- and D-strings, the Type II
string coupling $g_{\rm s}$ enters because their tensions differ by
the factor of $g_{\rm s}$.
Comparing Eqs. (\ref{magslope}) and (\ref{chargeslope}), to have
the F-D bound state we require $g_{\rm s} =1$. This does not ruin
perturbertivity because coupling constant is also inversely
proportional to volume, as in (\ref{4dcoupling}), which can vary.
For the higher dimensional states where the BPS condition is
saturated only by D-branes, we do not have such a $g_{\rm s}$
dependence. In fact, this system is converted into a bound state of
D0-D2 with the same BPS condition (\ref{survsusy}), by $U$-duality
transformations and $g_{\rm s}=1$ corresponds to the self-dual
coupling.

The preserved supersymmetry is identical when we have F-strings
along $X^1$,
\begin{equation}
 \epsilon = - \Gamma^0 \Gamma^2 \epsilon,
\end{equation}
which is identical to (\ref{survsusy}). In addition, we have the
D-brane along $X^2$ then
\begin{equation}
 \tilde \epsilon = \Gamma^0 \Gamma^1 \epsilon
\end{equation}
where $\tilde \epsilon$ is another supersymmetry parameter of the
opposite chirality to $\epsilon$. This breaks additional half of the
supersymmetries, generated by linear combination of left and right
chiral supercharges $\epsilon Q + \tilde \epsilon \tilde Q $. The
preserved supersymmetries are 1/4 of the original supersymmetries
and the same solution of (\ref{BPS}) is {\it preserved no matter how
the intersection is deformed}.

\begin{figure}[t]
\begin{center}
\epsfig{file=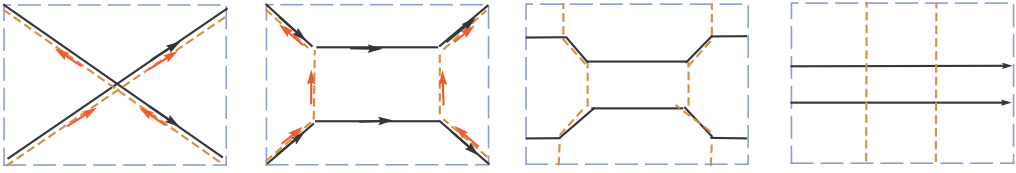,height=2.7 cm}
\end{center}
\caption{Transition of intersecting branes. Thick lines denote
D-branes and dotted lines denote electric fluxes. Brane and charge
orientations are denoted by arrows.  Every step in this figure is on
the same flat potential, therefore irreversible. }
\label{reversible}
\end{figure}

In Fig. \ref{reversible} we show  a simple but a typical situation
for deformation of the intersecting branes. At every state the same
supersymmetry is preserved and eventually the intersecting branes
are always deformed into parallel branes without costing energy. In
the last figure, we have parallel branes (with F-string winding).
The studies in Refs. \cite{AHK} have shown that various intermediate
state corresponds to a distinct phase due to interplay between
monopoles and instantons.

When we consider Chern--Simons couplings as well \cite{DM}, we have
a more general bending. We have an object that does not ruin
supersymmetry condition (\ref{survsusy}), the D$(-1)$brane, charged
under R--R scalar. We can see the deformation as distribution of
melted F-strings or electric field in the D-strings \cite{HT}.

\subsection{Higher dimensions}

The above (2+1) dimensional example is understood as a dimensional
reduction of ($d$+1) dimensional one. At first glance, this is not
possible because there seems no ($d$+1) dimensional objects carrying
F-string (Neveu--Schwarz--Neveu--Schwarz (NS--NS)) charge (except
NS5 brane). However there have been arguments \cite{LR1,LR2} that
continuous F-string filling in higher dimension can do such a job,
called (F, D$p$) bound state. The direct argument goes as follows.
Consider the above F1-D1 bound state. When we compactify one
transverse direction and take its $T$-dual, we obtain F1-D2 bound
state, where F1 string is melt to D2 brane. The fact of the
`boundedness' state is not affected in the $T$-dual picture as in
the D$p$-D$p'$ bound state and also it can be verified by checking
the number of supersymmetries. The shape of smaller brane is fuzzy
and totally melt and the dimension of F-string is not important.
Using $T$-duality iteratively in the transverse
directions, we easily obtain an F1-D$p$ bound state. It has been
argued in Ref. \cite{LR1} and an explicit supergravity solution is
obtained in Ref. \cite{LR2} so that a supersymmetry condition
similar to (\ref{balance}) is applied for such extended objects,
\begin{equation}
 T_{p,(m,n)} = T_p \sqrt{m^2 + {\frac {n^2} {g_{\rm s}^2}}}
\end{equation}
where $T_p = (4 \pi ^2 \alpha')^{(11-p)/2}$ is the pure D$p$ brane
tension \cite{Po}. Being the extension of F-string, it
continuously smears out the transverse direction with density
$T_p$, filling $(p-1)+1$ dimensional plane.  Although it fills
higher dimensional plane, it still remains as an F1 string and
this fact evades the contradiction due to the fact that there is
no higher rank NS--NS field other than $B_{\mu \nu}$.  We can
check the brane junction conditions (\ref{chargeslope},
\ref{BPSnet}), in particular relations between $p$ and $q$ are the
same. We have a direct dimensional reduction. This solution is
valid only for {\it bound states}, which is different from the
NS5-D5 nonbound states, for example.

Can we consider a more general solution, where the situation cannot
be projected into (1+2) dimensions? This corresponds to a typical
intersecting brane model, where branes are wrapping 3-cycles in six
torus $T^6$. We have an affirmative answer because in each
two-subtorus we can always satisfy the junction conditions
(\ref{chargeslope},\ref{BPSnet}) for arbitrary deformed D-branes,
because the supersymmetry condition (\ref{survsusy})  holds
regardless of the deformation of D-branes. It is also noted that the
BPS-Laplace equation allows a linearly growing solution. In fact, in
the case of noncompact dimensions, the linearly growing solution is
a good and normalizable one \cite{AHK}.

\subsection{Charge conservation and selection rules}

\begin{figure}[t] \begin{center}
\epsfig{file=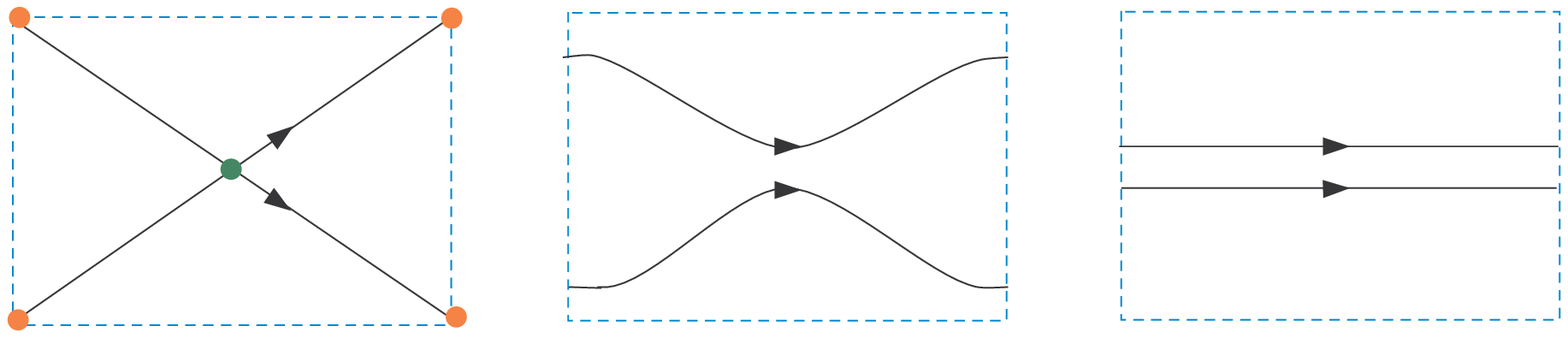,height=3 cm}
 \caption{Recombination of
branes on the torus. The transition $U(2) \leftrightarrow U(1)
\times U(1)$ occurrs. The parallel case can be either $U(2)$ or
$U(1)\times U(1)$ depending on the separation. Dots with the
identical color in the initial state denote the same intersecting
point. Arrows indicate the orientation of branes, or the
Ramond--Ramond charges. The process is reversible because the
geometry is determined by the D-flat moduli.} \label{parallel}
\epsfig{file=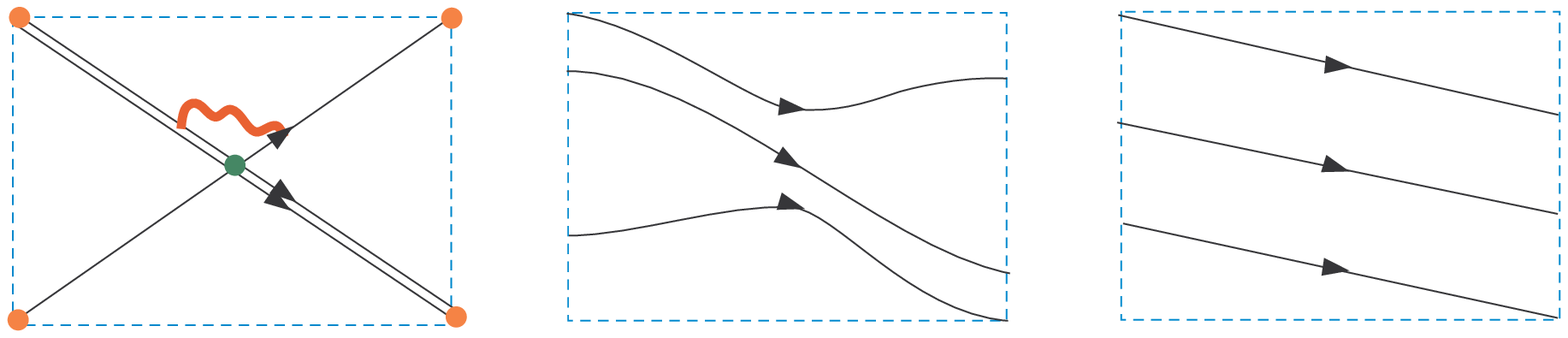,height=3 cm}
 \caption{A more
nontrivial setup compared to the preceeding figure. The intersecting
case has the $U(2) \times U(1)$ gauge fields and a chiral fermion
$\bf (2,1)$ localized at the intersection. We cannot make them
coincident because the winding number along the vertical direction
is preserved. With more branes forming stacks, we can explain the
number of families. Alternatively, in the presence of orientifold
plane, forced by tadpole cancelation, the winding number is summed
up to zero corresponding to the branes coincident into one stack.}
\label{comprecomb}
\end{center}
\end{figure}

It is important to note that {\it R--R and NS--NS charges must be
conserved,} unless a D-brane is created nor annihilated. Any brane
recombination takes place as long as each winding number is
conserved
\begin{equation}
 \sum N_a (n_a, m_a) = \text{conserved.}
\end{equation}
 Fig. \ref{parallel} shows the typical
features, where the sum of the charges in one direction vanishes,
for example $(n,m)=(2,0)$. After the brane separation, the two
slices of branes can describe either $U(2)$ or $U(1) \times U(1)$
depending on the separation, controlled by the (matrix-valued) VEV
of modulus $ X=2\pi \alpha' A$. This gives the selection rules for
the brane recombination process. We can freely move branes,
guaranteed by the BPS saturated supersymmetry, in the way that {\it
the R--R charge is conserved}. This is also interpreted as the
conservation of the brane winding number. We know that, when the
strings  merge or separate, the winding numbers are conserved in
each direction of the torus.

This feature is shown in Figs. \ref{parallel} and \ref{comprecomb}.
Because we make a torus identifying sides of a rectangle, we have
another intersection point on the edge. In Fig. \ref{parallel}, the
R--R charge or the winding numbers of the initial state is $(1,-1)$
and $(1,1)$ and that of the final state is $(2,0)$. The sum is
preserved. The final state is parallel and is made coincident. The
moduli space point corresponding to this coincident point allows the
gauge enhancement to $U(2)$. A nontrivial case is the example shown
in Fig. \ref{comprecomb} where the sum relation is
$2(1,-1)+(1,1)=(3,-1)$. In this case the three branes cannot move
into a coincident point to one stack, because the winding number in
the vertical direction is preserved with $n=-1$. A $U(3)$
enhancement does not occur. However, one can use the fact that the
Lagrangian describing this situation is still a broken phase of
$U(3)$.

\section{Toward top-down approach}

\subsection{Gauge coupling unification}

So far, we have discussed that there is a supersymmetry preserving
deformation making all the branes parallel. Of course, the
parallel and coincident brane states are described by the unique
$U(n)$ gauge group and the gauge couplings are unified. (In the
next subsection, we will consider the R--R tadpole cancelation
condition requiring negatively charged objects. When the typical
solution of orientifold is introduced, we always end up with a
situation with all the branes coincident.)

Note that the above vacua are supersymmetric and hence there {\it
needs no cost} on deformation because they are in $D$-flat
directions. It would be rather strange if the resulting {\it four
dimensional} couplings, dependent on compact volume factor $V_p$ as
in (\ref{4dcoupling}), {\it differ} between the parallel and
the intersecting case (and intermediate steps). The puzzle arises because
the naive expectation on volume dependence ignores the electric flux
dependence which is not interpreted geometrically.

The desired form is read from the fluctuation spectrum analysis
\cite{AHT}. That is, we expand the gauge field $A_\mu$ around the
intersecting brane background $A^0_\mu$,
\begin{equation}
 A_\mu = A^0_\mu + \delta A_\mu.
\end{equation}
Plugging it into the DBI action (\ref{DBI}) and expanding up to
the quadratic order in $\tilde F_{\mu \nu} \equiv \partial_{[\mu}
\delta A_{\nu]}$, around
\begin{equation}
 K{_\mu}^\nu = (1 + 2 \pi \alpha' F^0){_\mu}^\nu,
\end{equation}
 we obtain
\begin{equation}
 {\cal L} = \sqrt{ - \det K_{\mu \nu} }
 \left( K^{-1}_{\mu \nu} \tilde F^{\nu
 \lambda} K^{-1}_{\lambda \sigma} \tilde F^{\sigma \mu} +
 \text{(topological terms)}\right).
\end{equation}
Here $F^0_{\mu \nu}=\partial_{[\mu} A^0_{\nu]}$ from background
branes and raising and lowering indices are done with the
`genuine' flat metric $\eta_{\mu \nu}$. The topological term does
not contribute to the gauge coupling, and hence
\begin{equation}
 {\cal L} = \sqrt{ - \det K_{\mu \nu} } \left( g^{\mu \nu} \tilde F_{\nu
 \lambda} g^{\lambda \sigma} \tilde F_{\sigma \mu} + \dots \right).
\end{equation}
Here the nonvanishing contribution of $K^{-1}$ is only the symmetric
part \cite{AHT}, which we call `metric' $g$. Note that the YM action
cannot catch this volume dependence on $\sqrt{-\det K}$ and $g$ in
(\ref{4dcoupling}) so the {\it full DBI action} is required.

When only the magnetic flux $f_{12}=f_0$ is turned on, we have
\begin{equation}
  K^{\mu \nu} = \begin{pmatrix}
 -1 & 0 & 0 \\ 0 & 1 & 2 \pi \alpha' f_0 \\ 0 & -2 \pi \alpha' f_0 & 1
 \end{pmatrix}.
\end{equation}
(We deal with $(2+1)$ dimensional case and the generalization is
straightforward.) Then we have
\begin{equation}
 g^{\mu \nu} = \diag \left(-1, {1 \over 1+(2 \pi \alpha' f_0)^2},
 {1 \over 1+(2 \pi \alpha' f_0)^2} \right),
\end{equation}
which takes into account of the correct spacing of energy of
spectrum when branes are tilted. We see that the overall factor
becomes
\begin{equation} \label{ovfactor}
\sqrt{1+(2\pi \alpha' f_0)^2} = \sqrt{1+\tan^2 \theta} = {1 \over
\cos \theta},
\end{equation}
where the second relation comes from the $T$-dual interpretation.
Magnetic flux becomes geometry and gives rise to the extra factor
$V_p = L_1 L_2 / \cos \theta$, which is nothing but the volume
dependence of tilted branes from geometry, in (\ref{4dcoupling}).
Note that the metric in the time direction is $g^{00}= -1$, thus it
does not affect the canonical normalization of coupling.

When we have electric flux turned on in addition, satisfying the
BPS condition $F_{01}=F_{12}=f_0$,  we have
\begin{equation}
 K = \begin{pmatrix}
 -1 & 2 \pi \alpha' f_0 & 0 \\ -2 \pi \alpha' f_0 & 1 & 2 \pi \alpha'
 f_0 \\ 0 & -2 \pi \alpha' f_0 & 1
 \end{pmatrix}
\end{equation}
Remarkably the overall factor in (\ref{ovfactor}), responsible for
normalization of coupling, becomes
\begin{equation}
 \sqrt {- \det K} = 1
\end{equation}
thus the volume dependence becomes to $V_p=1$, which is the same
as the parallel brane case. {\it Whatever shape the branes take,
we have the same four dimensional gauge coupling; the gauge
couplings are unified.}

Of course, this modification is due to the effect of electric flux
$F_{01}$. One may worry that there might be extra factor in $g^{00}$
which ruins the canonical normalization. However, when we plug in
the new form of $K^{-1}$ and diagonalize it, the symmetric part in
the `metric' becomes
\begin{equation}
 g^{\mu \nu}_{\diag} = \diag \left(-1,
 {1 \over 1+2(2 \pi \alpha' f_0)^2},
 {1 \over 1+2(2 \pi \alpha' f_0)^2} \right).
\end{equation}
confirming $g^{00}_{\diag}=-1$. The nonabelian case shows
essentially the same feature, although it cannot reproduce the
correct infrared spectrum in the intersecting brane case.

\subsection{Including orientifold}

\begin{figure}[t]
\begin{center}
\epsfig{file=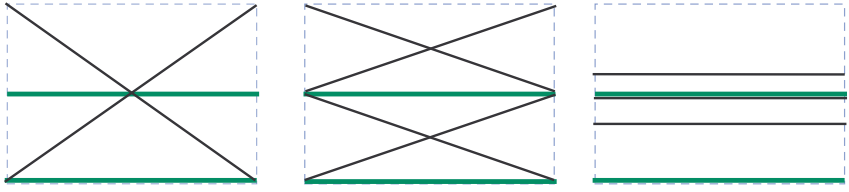,height=3 cm}
\end{center}
\caption{With orientifold the R--R tadpole cancelation condition
is much more restricted. By the reflection symmetry, the sum of
winding numbers transverse to the O-plane should vanish.}
\label{oplanepresent}
\end{figure}

In open string theories, we need the Ramond--Ramond tadpole
cancelation as a consistency condition for equation of motion
\cite{RZ}. The condition is derived from various 1-loop diagrams of
open strings with Euler number one. Since it is dependent on the
D-brane charge which winds topological cycles, it is converted to
the charge conservation in the compact space. In our case of $(2+1)$
dimensional setup, the cancelation condition reads
\begin{equation} \label{RRcancel2d}
 [\pi_{\rm total}] = \sum N_a (m_a,n_a) = 0.
\end{equation}
For example, in Fig. \ref{intbrane}, the winding number is $(1,1)$
for $U(3)$ stack and $(1,2)$ for $U(2)$ stack. We can verify that
the sum of charge is not zero. To cancel the tadpole, we should
insert a D-brane like object having the opposite R--R charge. The
best choice is the orientifold plane (O$p$ plane). Its tension is
\begin{equation}
 T_{{\rm O}p} = \mp 2^{p-5} T_{{\rm D}p}\ ,
\end{equation}
where the signs taking account of relative orientation with respect
to D-branes. Therefore in effect we have a charged object of$(\mp
2^{p-5},0)$. Because of the reflection symmetry, the sum of R--R
charges is zero.

In addition, we should always have the `mirror brane' with respect
to the orientifold plane, as shown in Fig. \ref{oplanepresent}. That
is, for every $(n,m)$ winding state, we have an $(n,-m)$ winding
state. Therefore, according to the selection rule we always have the
state with a single stack of D-branes, aligned parallel to the
direction of the orientifold plane. Since all are parallel, we can
make them all coincident with orientifold planes. The final state is
a gauge theory with $SO(2n)$.

The dependence of $2^p$ is interpreted as follows. Because the
orientifold is related by $\Z_2$ orbifold fixed points, we have
twice the number of $\Z_2$ fixed points when we compactify a
dimension. For $p=9$, for the space-filling orientifold the charge
is 32 times that of the D-brane. This is nothing but the setup we
obtain Type I theory out of Type IIB theory, by moding out the
chirality. {\it It strongly suggests that therefore many models
are obtained from Type I theory with SO(32) gauge group.}

Even if we include orientifold planes in various dimensions, the
meaningful setup arises as $T$-dual of Type I string with O5 plane
(also and/or O1 planes), because to preserve the number of
supersymmetry we require the number of Neumann--Dirichlet
direction (which is invariant quantity under $T$-dual) to be a
multiple of four.

Of course, all the setups of intersecting branes at angles are not
obtained in this way, because the final intersecting states
considered here have more supersymmetries than ${\cal N}$=1.
Nevertheless, we observe the following lesson learnt from the
traditional compactification of heterotic strings. First, many
supersymmetric models rely on the same supersymmetries, namely the
small number of supersymmetries are obtained not by the unstable
setup of intersection but the compactificaton on manifold with a
certain holonomy. The realistic four supersymmetries (${\cal N}=1$
in four dimensions) is obtained by the orbifold projection
\cite{DHVW} such as $T^6/\Z_N$ or $T^6/\Z_M \times \Z_N$ \cite{CSU},
but with the explicit supersymmetry breaking setup. Also, we have
experienced that many vacua of heterotic string \cite{Ch,CGT} is
connected when we consider $(2,2)$ compactification \cite{Ib89}. We
may conjecture that a vacuum of less supersymmetric setup is also
connected to a more symmetric setup.

\section{Conclusions}
In this paper we studied a supersymmetry preserving deformation of
branes. The symmetry of nonabelian gauge group is understood as
parallel branes. Their separation gives rise to gauge symmetry
breaking and can be interpreted as spontaneous symmetry breaking
along flat directions. With more general deformations we introduce a
deeper understanding on the symmetry breaking. We have the selection
rule for charge conservation and showed the possible deformation to
the state where all branes are parallel. It is described by a unique
gauge group and suggests the unification of gauge coupling
constants.

Supersymmetry is maintained by turning on a suitable electric flux,
which is interpreted as a F-string bound state or string junction.
We can generalize this idea to higher dimensions to form a (F, D$p$)
bound state and junction. Since this process does not ruin
supersymmetry, we naturally maintain the supersymmetry at the high
energy scale when we consider this process. This is due to the
interplay between electric and magnetic flux.

Remarkably, in this situation, we have shown that at every step, we
have the same four dimensional gauge couplings, contrary to the
conventional case when we have only magnetic flux. Thus we achieve
gauge and coupling unification.

To cancel the Ramond--Ramond tadpole amplitude, we introduce
orientifold planes. We have then the transition to the state where
{\it totally parallel and coincident branes} along some
coordinates and this is thought as some $T$-dual state of Type I
string theory having gauge group $SO(32)$.

Here, we discussed only the cases for toroidal compactification and
have not done the orbifold cases. However, we might conjecture that,
like in the heterotic string theory, the realistic models seem to
originate from a single theory. In this scenario, supersymmetry
breaking is done by compactification on Calabi--Yau manifolds or
orbifolds. In this sense, this paper provides a relevant scenario of
top-down approach in the brane setup.

\acknowledgments

KSC is grateful to Daniel Cremades, Seungjoon Hyun, Hyung Do Kim,
and Michele Trapletti for useful discussions and especially to Koji
Hashimoto for motivation and answers to important questions. This
work is supported in part by the KRF Sundo Grant No.
R02-2004-000-10149-0, the KRF ABRL Grant No. R14-2003-012-01001-0
and the BK21 program of Ministry of Education.

\end{document}